# Efficient spin torques in antiferromagnetic CoO/Pt quantified by comparing field- and current- induced switching

L. Baldrati[1*], C. Schmitt[1], O. Gomonay[1], R. Lebrun[1,2], R. Ramos[3], E. Saitoh[3,4,5,6,7], J. Sinova[1,8,9], M. Kläui[1,9*]

*[1]Institute of Physics, Johannes Gutenberg-University Mainz, 55128 Mainz, Germany*

*[2]Unité Mixte de Physique CNRS, Thales, Univ. Paris-Sud, Université Paris-Saclay, Palaiseau 91767, France*

*[3]WPI-Advanced Institute for Materials Research, Tohoku University, Sendai 980-8577, Japan*

*[4]Institute for Materials Research, Tohoku University, Sendai 980-8577, Japan*

*[5]Advanced Science Research Center, Japan Atomic Energy Agency, Tokai 319-1195, Japan*

*[6]Center for Spintronics Research Network, Tohoku University, Sendai 980-8577, Japan*

*[7]Department of Applied Physics, The University of Tokyo, Tokyo 113-8656, Japan*

*[8]Institute of Physics, Academy of Sciences of the Czech Republic, Praha 11720, Czech Republic*

*[9]Graduate School of Excellence Materials Science in Mainz, 55128 Mainz, Germany*

*\*Electronic Mail: lbaldrat@uni-mainz.de, Klaeui@Uni-Mainz.de*

## ABSTRACT

We achieve current-induced switching in collinear insulating antiferromagnetic CoO/Pt, with fourfold in-plane magnetic anisotropy. This is measured electrically by spin Hall magnetoresistance and confirmed by the magnetic field-induced spin-flop transition of the CoO layer. By applying current pulses and magnetic fields, we quantify the efficiency of the acting current-induced torques and estimate a current-field equivalence ratio of $4 \times 10^{-11}$ T A$^{-1}$ m$^2$. The Néel vector final state ($\boldsymbol{n} \perp \boldsymbol{j}$) is in line with a thermomagnetoelastic switching mechanism for a negative magnetoelastic constant of the CoO.

# MANUSCRIPT

Antiferromagnetic materials (AFMs) are considered important future materials for spintronics, thanks to advantageous properties compared to ferromagnets, that potentially enable higher speed (resonance frequencies in the teraHertz range), bit packing density (absence of generated stray field) and resilience to external applied magnetic fields [1]. However, exploiting AFMs in applications requires electrical reading and writing of information, which can be stored e.g. in the orientation of the antiferromagnetic Néel vector $\boldsymbol{n}$. Recently, this has been reported by electrical measurements and direct magnetic imaging both in metallic AFMs [2,3] and bilayers of insulating AFMs and heavy metals [4–9]. The underlying switching mechanism in the latter case is being debated, in terms of both origin and efficiency [4–7]. While different claims have been made, a key missing step is the experimental quantification of the acting torques in compensated AFMs, which enables comparison to future *ab initio* calculations. This has been prevented so far, by the difficult reading of the antiferromagnetic state, the presence of electrical signal artefacts not related to the antiferromagnetic order [6,7,10–12] and the difficulties in controlling the orientation of $\boldsymbol{n}$ by an external magnetic field $\boldsymbol{H}$. To quantify the torques, one needs to study compensated AFMs with low anisotropy that present an accessible spin-flop transition, i.e. the reorientation from $\boldsymbol{n} \parallel \boldsymbol{H}$ to $\boldsymbol{n} \perp \boldsymbol{H}$.

A possible material with apt properties is CoO, a collinear compensated antiferromagnet with Néel temperature $T_{Néel} = 291\ K$ in the bulk [13–15], and spin-flop transition at $12\ T$ and $77\ K$ [16]. By growing CoO thin films under a compressive strain on MgO (lattice mismatch 1.1%) [17,18], one can induce an in-plane easy magnetic configuration and $T_{Néel}$ around room temperature. In MgO/CoO/Fe thin films it was conjectured, by looking at the Fe anisotropy, that the CoO layer has fourfold in-plane anisotropy [19]. The existence of a spin-flop transition for such strained thin films with in-plane easy axes has not been investigated, but, if accessible, may prove suitable to compare current- and field-induced switching efficiencies quantitatively.

In this letter, we quantify the torques due to current injection in the CoO/Pt system. First, we show that the compressive strain favors a fourfold in-plane magnetic anisotropy of the CoO layer with two easy axes in the (001) plane. Having two orthogonal stable states is ideal for applications where the orientation of $\boldsymbol{n}$ is read by spin Hall magnetoresistance (SMR) [5,20,21]. Second, we achieve electrical switching and probe its symmetry, showing that this switching is of magnetic origin and not related to the Seebeck effect [11] or to electromigration effects that we identify for particular conditions as well [7,10]. Finally, we directly compare the effects of the field and current pulses in a Pt layer on the reorientation of $\boldsymbol{n}$ in the CoO, quantifying the current-field equivalence of the current-induced torques, showing that currents are much more efficient than magnetic fields for the switching of AFMs.

After optimizing the epitaxial CoO/Pt thin film growth [22–24], we first probe electrically the magnetic anisotropy of the CoO by means of uniaxial field-sweep magnetoresistive scans (MR) and angularly detected magnetoresistance scans (ADMR) in patterned Hall bar devices oriented along the [100] direction [20,21,25,26]. The electrical measurements were performed in a cryostat, equipped with a variable temperature insert, a rotating sample stage and a superconducting magnet generating fields up to 12 T. The orientation of $\boldsymbol{n}$ can be read electrically, by means of the transverse SMR signal, proportional to the in-plane Néel vector components $n_x * n_y$, according to the geometry shown in Fig. 1a,b. Note that the SMR is maximized when two states with orthogonal orientation of $\boldsymbol{n}$ are present in the system. The resistance was measured by a Keithley 2400 and a Keithley 2182 and averaged between opposite DC current polarities of density $j_{meas} \sim 5 \times 10^9\ A\ m^{-2}$, thus minimizing thermally-induced electric effects, similar to the protocol developed for antiferromagnetic hematite [27]. When the field is applied alternated along the [110] or [$\bar{1}$10] directions (easy axes) at 8 T and 200 K, we find an abrupt spin-flop transition in a MgO(001)//CoO(5 nm)/Pt(2 nm) sample, as shown in Fig. 1c,d. The resistance change at the spin-flop is consistent with a negative sign of the SMR [20,21,25,26]. Moreover, applying a field along the [001] out-of-plane direction (hard axis) does not lead to a spin-flop below 12 T at 200 K, in line with a biaxial in plane magnetic anisotropy. We did not find a hysteresis loop in the MR, showing that the CoO(001) interface is likely fully compensated [14]. By looking at the ADMR in Fig. 1e, one can see a $\sin^2(\alpha + \alpha_0)$ signal [23] and three distinct hysteresis loops, centered around the $\alpha = 0°$ [100], $\alpha = 90°$ [010] and $\alpha = 180°$ [$\bar{1}$00] directions (hard axes), while the resistance is not hysteretic around the $\alpha = 45°, \alpha = 135°$ (easy axes). The hysteresis loops, according to a macrospin model (Supplementary Ref. [22]), are due to the lag of $\boldsymbol{n}$ behind the rotation of $\boldsymbol{H}$ in the vicinity of the hard axes (HAs), while we observe field-induced spin-flop of $\boldsymbol{n}$ in the vicinity of the two orthogonal in-plane easy axes (EAs). These observations demonstrate the fourfold in-plane magnetic anisotropy of the CoO layer induced by the strain, with an out-of-plane hard axis along the [001] direction and two easy axis in the (001) plane ([110] and [$\bar{1}$10]), in agreement with the symmetry of the anisotropy conjectured in exchange-biased CoO/Fe thin films [19]. Moreover, we show in Fig. 1f that the spin-flop field vanishes at $T_{Néel} = 305 \pm 5\ K$. This is increased by 10 K compared to the bulk due to strain [17], in line with the literature.

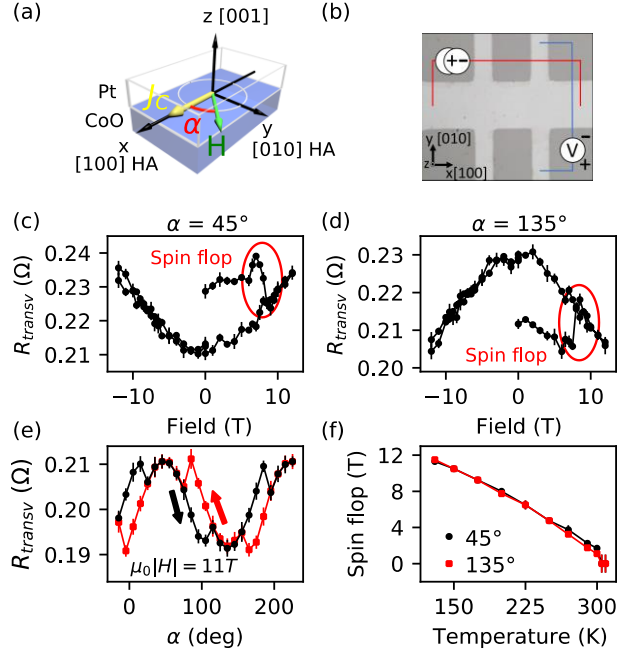

*Fig. 1: Magnetic anisotropy of the CoO thin films. (a) Coordinate system. (b) Optical micrograph of one Hall bar and contact scheme. (c) Field induced spin-flop read by SMR in the presence of a field applied along the [110] direction (α = 45°) in CoO(5 nm)/Pt(2 nm). A 12 T field was previously applied along the orthogonal direction. (d) Field-induced spin-flop with orthogonal field direction compared to the previous one. (e) ADMR transverse resistance measurements showing hysteresis loops associated with the spin-flop transition. (f) Spin-flop field versus temperature, yielding $T_{Néel} = 305 \pm 5\ K$.*

Next we need to ascertain that we can obtain current-induced switching in the fourfold CoO thin films. We use 8-arms Hall stars devices with the pulsing arms oriented along the [110] and [$\bar{1}$10] easy axes directions (Fig. 2a-d) at 200 K. To set a well-defined starting state, we applied $\mu_0 H_{before} = 11\ T$ along the [$\bar{1}$10] direction, i.e. along the 4-1 contacts as defined in Fig. 2a, and then reduced the field to $0\ T$, thus aligning before each pulse $\boldsymbol{n} \perp \boldsymbol{H}$ in the in-plane direction of the 3-2 contacts ([110]). In the case of Fig. 2a we applied 5 pulses 1 ms long and of current density $j_{pulse} = 1.15\ \text{x}10^{12}\ A\ m^{-2}$ along 3-2 (initial state $\boldsymbol{n} \parallel \boldsymbol{j}_{pulse}$) by a Keithley 6221, i.e. the pulses were applied with $\boldsymbol{j}_{pulse} \perp \boldsymbol{H}_{before}$. The transverse resistance, measured 10 s after the application of the pulses, drops after the first pulse, in a step-like fashion that was also reported in NiO/Pt [4,6], indicating a current-induced 90° $\boldsymbol{n}$ rotation analogous to the spin-flop transition. If one performs a MR scan with field along 4-1 after the current pulses, shown in Fig. 2b, one observes a field-induced spin-flop transition of $\boldsymbol{n}$ back to the initial state (along [110]). Note that the height of the current-induced switching in Fig. 2a (red-arrow) and of the field-induced spin-flop in Fig. 2b (red-arrow) are identical within the error and have the same magnitude as the spin-flop induced by a field only (Supplementary Ref. [22]), suggesting that both fields and currents switch $\boldsymbol{n}$ in the same manner. From the presence of a spin-flop after the 3-2 current pulse (Fig. 2b), considering that a spin-flop occurs only when $\boldsymbol{H} \parallel \boldsymbol{n}$, we determine that the switching final state is $\boldsymbol{n} \perp \boldsymbol{j}_{pulse}$. Accordingly, if after applying a field along [$\bar{1}$10], five current pulses are applied along the same direction 1-4 [1$\bar{1}$0], no transverse resistance variation and subsequent spin-flop transition is seen in the field scan (Fig. 2c,d), as in this case the initial state $\boldsymbol{n} \perp \boldsymbol{j}_{pulse}$ is already coincident with the final state observed after a current pulse. The switching can be reversed by sending current pulses in alternating orthogonal pulsing arms of the device, in the absence of any field. We show in Supplementary Ref. [22] approximately 350 current-induced switching events, without breaking the device. The current pulse polarity does not play a detectable role for the switching. These results confirm unambiguously the electrical reading and writing of the orientation of $\boldsymbol{n}$ in AFMs.

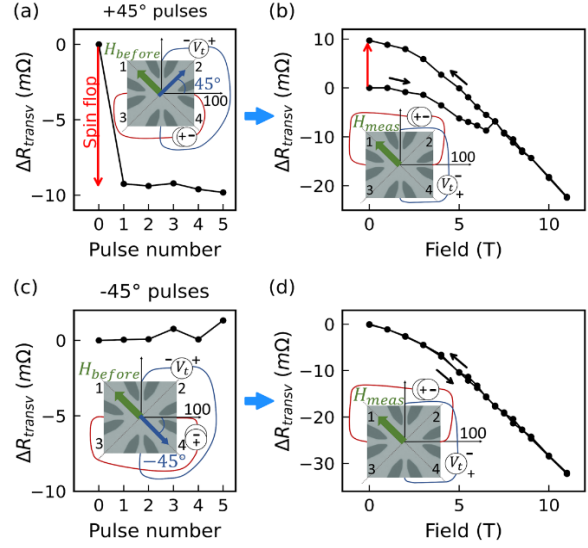

*Fig. 2: Symmetry of the current-induced switching. (a) $\mu_0 H_{before} = 11\,T$ was applied along the 4-1 contacts direction $[\bar{1}10]$ to align the Néel vector $\mathbf{n}$ along 3-2 $[110]$ and then removed. A step-like switching by pulses along 3-2 (starting state $\mathbf{j}_{pulse} \parallel \mathbf{n}$) is seen, corresponding to a current-induced spin-flop transition of $\mathbf{n}$ along 4-1 (final state $\mathbf{j}_{pulse} \perp \mathbf{n}$). (b) The MR measurement with field along 4-1 shows a field-induced spin-flop, which resets $\mathbf{n}$ along 3-2. (c, d) No switching and spin-flop are observed by pulses $\mathbf{j}_{pulse} \perp \mathbf{n}$, as this is already the final state.*

Finally, to quantify the current-field equivalence in the CoO(5 nm)/Pt system, we study the current-induced switching in the presence of static magnetic fields, applied along or perpendicular to the initial $\mathbf{n}$ of the system, during the current pulse. In Fig. 3a we show an example of this type of measurements for a single field, where we prepare the system in the same reproducible starting state with a reset pulse along 3-2 of $j_{reset} = 1.05 \times 10^{12}\,A\,m^{-2}$ and vary $j_{pulse}$ of the subsequent pulses along 1-4, as shown in Fig. 3b. By the saturation level of the transverse resistance we can determine the switching fraction assuming it is proportional to the resistance increase and equal to 100% at saturation. The amplitude of the switching as a function of pulse current and field is shown in Fig. 3c, where the color indicates the switching fraction (the darker the higher). The main result is that both the threshold and the saturation current are increased (decreased) if the field is applied orthogonal (parallel) to the initial orientation of $\mathbf{n}$. This is consistent with the fact that the Zeeman energy is minimum in antiferromagnets when $\mathbf{H} \perp \mathbf{n}$ [21]. By interpolation of the data, we can obtain the contour plots of equal switching efficiency that can be fitted by linear functions having $R^2$ values larger than 0.87, thus indicating that a linear relation between the field and the current can explain well the data. From the fits and considering the geometry of the device, we obtain a current-field equivalence of $4 \times 10^{-11}\,T\,A^{-1}\,m^2$, several orders of magnitude larger than the value $10^{-15}\,T\,A^{-1}\,m^2$ obtained in typical ferro(i)magnetic insulators, such as TIG/Pt [28]. The switching current density at zero field in CoO/Pt is $j_{pulse} = 6.5 \times 10^{11}\,A\,m^{-2}$ for a switching fraction of 15% and $j_{pulse} = 8.5 \times 10^{11}\,A\,m^{-2}$ to achieve a full switching, similar to what is found in TIG/Pt [29]. This shows that the obtained giant current-field equivalence ratio results from a current-induced switching that is equally efficient as in ferro(i)magnetic insulators, while the field-induced switching is very inefficient due to the insensitivity of AFMs against external magnetic fields. Note that we find a similar order of magnitude if we use a second method to estimate the current-field equivalence, namely by switching with pulses of increasing current density and looking at the increasing spin-flop field of the switched states (Supplementary Ref. [22]).

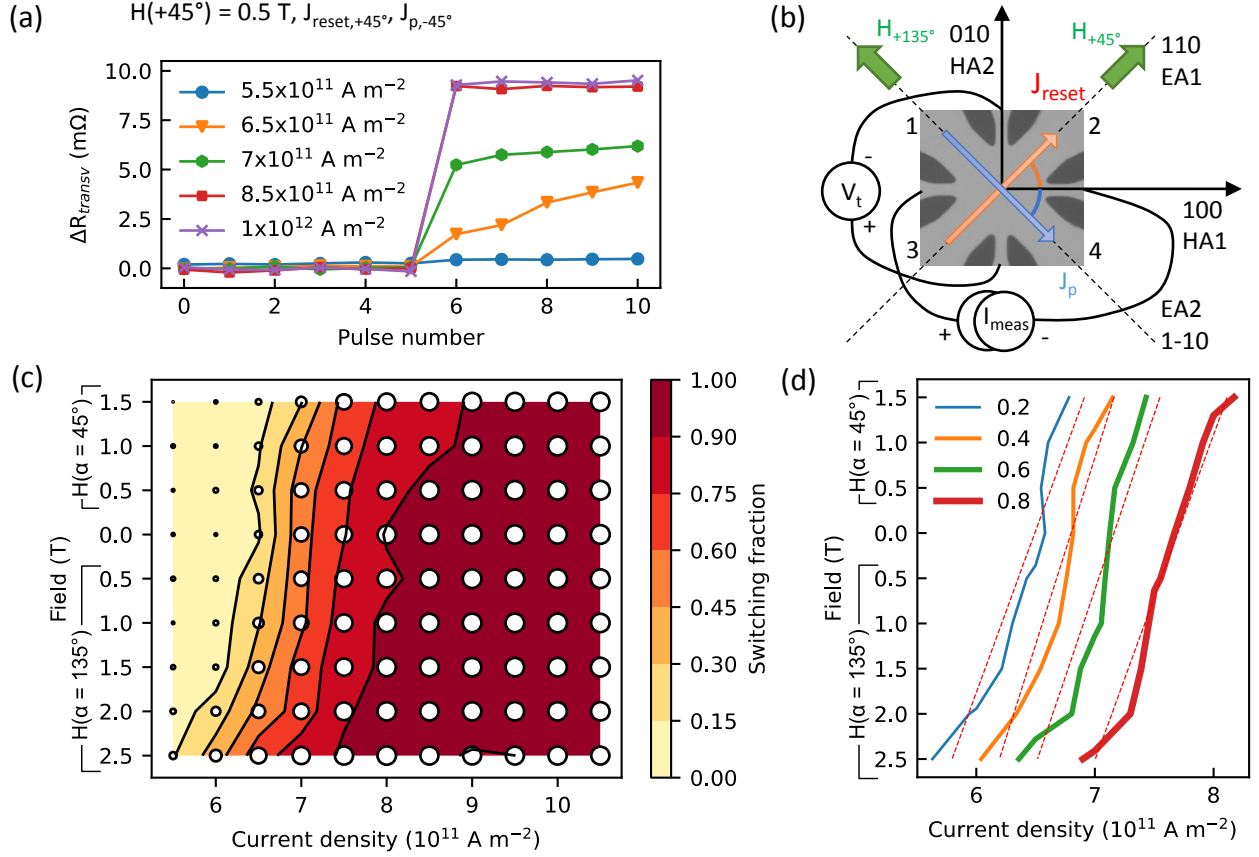

*Fig. 3: (a) Transverse resistance variation versus pulse current density, probing the threshold and saturation of the switching. Before the measurements a reset pulse ($j_{reset} = 1.05 * 10^{12}\ A\ m^{-2}$) was applied along 3-2, followed by 5 pulses along 3-2 and 5 pulses along 1-4. (b) Scheme of the measurements. (c) Switching fraction as a function of the applied field and pulse current. The circles represents the data points, the lines are contour plots with constant switching efficiency. (d) Current-field equivalence obtained by linear fits of the contour plots from the data in Fig. 3c for different switching fractions.*

To understand the field-current equivalence of the torques and the occurring switching mechanism, we consider the different torque mechanisms proposed to date: the damping-like spin-orbit torques (SOTs) acting on uncompensated ferromagnetic spins [4], the damping-like SOTs acting on the antiferromagnetic sublattices [5,6] and the thermomagnetoelastic effects [7]. The spin orbit torque (SOT) switching mechanism related to uncompensated interfacial spins [4] cancels out in our sputtered films with CoO(001) surfaces with a compensated checkerboard alignment of the spins [14] and the expected roughness and atomic steps. The mechanism based on SOTs in AFMs is related to the spin accumulation induced by the spin Hall effect [5]. The corresponding SOTs create staggered fields, which remove the degeneracy between the two orthogonal orientations of $\boldsymbol{n}$, leading to a current-induced energy term competing with the magnetic anisotropy [6]:

$$w_{\text{SOT}} = -\frac{\varepsilon^2}{H_\parallel M_s}(\boldsymbol{n}\cdot\boldsymbol{j}_{pulse})^2, (1)$$

where $H_\parallel > 0$ is the out-of-plane magnetic anisotropy of CoO, $\varepsilon$ is a material-dependent constant which parametrizes the coupling between the spin-current and the localized moments of the CoO layer. To minimize this contribution, the predicted final state of the switching is $\boldsymbol{n} \parallel \boldsymbol{j}_{pulse}$, opposite to what we probe. By comparing the expression (1) to the effective Zeeman energy contribution of the magnetic field

$$w_{Zee} = H_j H_k (n_j n_k - \delta_{jk})/H_{ex}, (2)$$

where $H_{ex}$ is the exchange field, we conclude that the spin-polarized current is linearly proportional to the effective magnetic field generated by the torque $\boldsymbol{H}_{SOT} \propto \hat{\boldsymbol{z}} \times \boldsymbol{j}_{pulse}$.

The third possible mechanism is related to Joule heating. It results from the combined effect of thermal expansion and magnetoelasticity [7]. According to this model, the degeneracy of the orthogonal states can be removed by the magnetoelastic contribution $w_{me} = \lambda u_{jk}^{rel} n_j n_k$ into the magnetic energy, where $\lambda$ is the magnetoelastic constant. The shear strains $u_{jk}^{rel}(\mathbf{r})$ compensate the stresses induced by the incompatibility of the thermal lattice (volume) expansion along the lines that separate high and low temperature regions. The strains along the direction of temperature gradient are tensile at the hotter side and are compressive at the colder side. In the center of the structure (where we read the SMR signal) the overall strain is compressive along the current direction (Supplementary Ref. [22]). The absolute value of the strain is proportional to the temperature gradient, but in general depends on the temperature distribution in the whole sample due to the nonlocality of the elastic interactions. However, as the temperature gradient is induced by Joule heating, $u_{jk}^{rel} \propto j_{pulse}^2$. Hence, in this region the current-induced contribution into the magnetic energy scales as

$$w_{me} \propto -\lambda (\mathbf{n} \cdot \mathbf{j}_{pulse})^2. \quad (3)$$

Assuming that the sign of the magnetoelastic constant in CoO is $\lambda < 0$ [30], the elongation in the direction of $\mathbf{n}$ is favored, which yields a final state $\mathbf{n} \perp \mathbf{j}_{pulse}$, resulting from the competition of pure magnetic and magnetoelastic anisotropies. Note that in general case the strains $u_{jk}^{rel}(\mathbf{r})$ depend on the distribution of the current density gradients with respect to the observation point and are not directly related with the direction of $\mathbf{j}_{pulse}$, in contrast to the case of SOTs. If we compare $w_{me}$ and $w_{Zee}$, one can see that the value of the effective magnetic field generated by thermomagnetoelastic effects is $H_{me} \propto j_{pulse}$ while its orientation is sensible to the geometry of the experiment and can be either parallel or perpendicular to the current direction.

Overall, both models predict a linear dependence of the effective field on the current density, as found experimentally. The final state after switching, found here in the discussion of Fig. 2 ($\mathbf{n} \perp \mathbf{j}_{pulse}$), is consistent with the final state expected from switching by the thermomagnetoelastic mechanism found in α-Fe$_2$O$_3$/Pt [7], and is opposite to the final state expected from switching due to an antiferromagnetic antidamping-like interfacial spin-orbit torque ($\mathbf{n} \parallel \mathbf{j}_{pulse}$) [5,6,9]. While both SOT and thermomagnetoelastic effects might be present, here the thermomagnetoelastic mechanism dominates. However, knowing the sign of the magnetostriction of CoO thin films is required to confirm that this mechanism leads to the observed final state of the switching, which has not been reported up to now in thin films. This thermomagnetoelastic mechanism can be stronger in CoO compared to other materials due to the large magnetostriction on the order of 10$^{-3}$ [31,32] and large out-of-plane magnetic anisotropy in our in-plane thin film samples, which can overcome the switching mechanism based on SOT effects in this material [6]. Also note that the combination of Eq. (2) and Eq. (3) explains the dependence on the field orientation that we found experimentally: when $\mathbf{H} \parallel \mathbf{j}_{pulse}$ ($\alpha = 135°$ in Fig. 3c,d) the two energy terms act constructively to decrease the current switching threshold, while when $\mathbf{H} \perp \mathbf{j}_{pulse}$ ($\alpha = 45°$), the current-switching threshold is increased (Supplementary Ref. [22]).

In conclusion, we report here the measured equivalence of current and field in antiferromagnetic CoO/heavy metal Pt bilayers, where the CoO is antiferromagnetic and has the fourfold in-plane anisotropy which is ideal for applications. First, our data clearly show that electrical reading and writing of the switching in antiferromagnetic materials is possible and achieved efficiently in CoO/Pt. Second, we find that the relation between current and field is linear and of magnitude much larger than in ferromagnets, with current-induced switching similarly efficient as in ferromagnets and the insensitivity of the AFMs against external magnetic fields. Third, the switching final state and current-field equivalence suggest that a switching mechanism based on thermomagnetoelastic effects is the likely origin of the observed switching.

**Acknowledgements**
The authors thank A. Ross, J. Henrizi, A. Dion, T. Reimer for skillful technical assistance. L.B. acknowledges the European Union's Horizon 2020 research and innovation program under the Marie Skłodowska-Curie grant agreement ARTES number 793159. O.G. acknowledges the EU FET Open RIA Grant no. 766566, the DFG (project SHARP 397322108) and that this work was funded by the Deutsche Forschungsgemeinschaft (DFG, German Research Foundation) – TRR 288 - 422213477 (project A09). L.B., R.L., and M.K. acknowledge support from the Graduate School of Excellence Materials Science in Mainz (MAINZ) DFG 266, the DAAD (Spintronics network, Project No. 57334897) and all groups from Mainz acknowledge that this work was Funded by the Deutsche Forschungsgemeinschaft (DFG, German Research Foundation) - TRR 173 – 268565370 (projects A01, A03, A11, B02, and B12). We acknowledge financial support from the Horizon 2020 Framework Programme of the European Commission under FET-Open grant agreement no. 863155 (s-Nebula). This project has received funding from the European Research Council (ERC) under the European Union's Horizon 2020 research and innovation programme under grant agreement 856538 (3D Magic). J.S. acknowledges support from the Grant Agency of the Czech Republic grant no. 19-28375X and ASPIN EU FET Open RIA Grant no. 766566. This work was also supported by ERATO "Spin Quantum Rectification Project" (Grant No. JPMJER1402) and the Grant-in-Aid for Scientific Research on Innovative Area, "Nano Spin Conversion Science" (Grant No. JP26103005), Grant-in-Aid for Scientific Research (S) (Grant No. JP19H05600), Grant-in-Aid for Scientific Research (C) (Grant No. JP20K05297), from JSPS KAKENHI, Japan. R.L. acknowledges the European Union's Horizon 2020 research and innovation program under the Marie Skłodowska-Curie grant agreement FAST number 752195.

# Efficient torques in antiferromagnetic systems quantified by comparing field- and current- induced switching of CoO/Pt thin film bilayers – supplementary information

L. Baldrati[1]*, C. Schmitt[1], O. Gomonay[1], R. Lebrun[1,2], R. Ramos[3], J. Sinova[1,8], E. Saitoh[3,4,5,6,7], M. Kläui[1,9*]

[1]*Institute of Physics, Johannes Gutenberg-University Mainz, 55128 Mainz, Germany*
[2]*Unité Mixte de Physique CNRS, Thales, Univ. Paris-Sud, Université Paris-Saclay, Palaiseau 91767, France*
[3]*WPI-Advanced Institute for Materials Research, Tohoku University, Sendai 980-8577, Japan*
[4]*Institute for Materials Research, Tohoku University, Sendai 980-8577, Japan*
[5]*Advanced Science Research Center, Japan Atomic Energy Agency, Tokai 319-1195, Japan*
[6]*Center for Spintronics Research Network, Tohoku University, Sendai 980-8577, Japan*
[7]*Department of Applied Physics, The University of Tokyo, Tokyo 113-8656, Japan*
[8]*Institute of Physics, Academy of Sciences of the Czech Republic, Praha 11720, Czech Republic*
[9]*Graduate School of Excellence Materials Science in Mainz, 55128 Mainz, Germany*
*Electronic Mail: lbaldrat@uni-mainz.de, Klaeui@Uni-Mainz.de*

## GROWTH OPTIMIZATION AND STRUCTURAL CHARACTERIZATION

CoO epitaxial thin films were grown by reactive magnetron sputtering using a ULVAC QAM 4 fully automated sputtering system. After pre-annealing the MgO(001) substrates at 770 °C for 2 hours, CoO was deposited from a Co target in a mixed Ar (15 sccm) and $O_2$ (2 sccm) atmosphere by RF magnetron sputtering at 430 °C and 150 W. The Pt top layer was subsequently deposited in a separate chamber of the same system after cooling down the sample to room temperature in vacuum. The growth conditions were checked by x-ray diffraction (XRD) and reciprocal space mapping measurements were performed with a Bruker D8 Discover high resolution diffractometer, with *Cu $K_\alpha$* radiation of wavelength equal to 0.15406 nm.

In Fig. S1a we show the XRD 2θ-ω scans of MgO(001)//CoO(t)/Pt(2) thin film samples for different thicknesses of the CoO layer. The peak position is compatible with the CoO(002) peak, thus indicating that the CoO orientation is (001) as the MgO substrate. The increasing 2θ value of the peak for increasing CoO thickness indicates a lower lattice constant for thicker films. Note that a CoO film 90 nm thick has a measured out-of-plane lattice constant $c = 0.4271 \pm 0.011$ nm, very close to the bulk value, while the value is increased in CoO 10 nm thick to $c = 0.4300 \pm 0.0121$ nm, and a 5 nm thick layer is expected to present an even larger *c* value due to the increased strain. This can be explained considering that CoO is grown under compressive strain, according to the lattice constants of bulk MgO (a = 0.4212 nm) and CoO (a = 0.4260 nm), lattice mismatch 1.1%. Due to the in-plane compressive strain, thin films have an out-of-plane lattice constant larger than in the bulk. The stress is gradually released for increasing CoO thickness.

In Fig. S1b,c we show symmetric and antisymmetric reciprocal space mapping (RSM) measurements at the 002 and 113 diffraction peaks of a MgO//CoO(25 nm)/Pt(2) sample. One can see that the CoO and MgO peak positions are aligned along almost the same *h* value, indicating that they have very similar in-plane lattice parameter, thus corroborating the cube-on-cube growth of the epitaxial CoO thin film layer on the MgO substrate, albeit the very slight deviation in the *h*-values indicates the presence of a small amount of relaxation.

The thickness of the layers was calibrated with x-ray reflectivity (XRR). By fitting the XRR curve of a MgO/CoO(25 nm)/Pt(2 nm) sample (not shown), we estimate the RMS roughness for both the CoO ($R_{q,CoO}$ ~ 0.6 nm) and the Pt ($R_{q,Pt}$ ~ 0.7 nm) layers. Moreover, we measured in a Hall bar device a resistivity of 2.8 x $10^{-7}$ Ω m at 300 K, close to the bulk value of 1 x $10^{-7}$ Ω m. The low Pt thin film roughness and the resistivity, similar to the bulk, together indicate that the Pt(2 nm) layer is continuous.

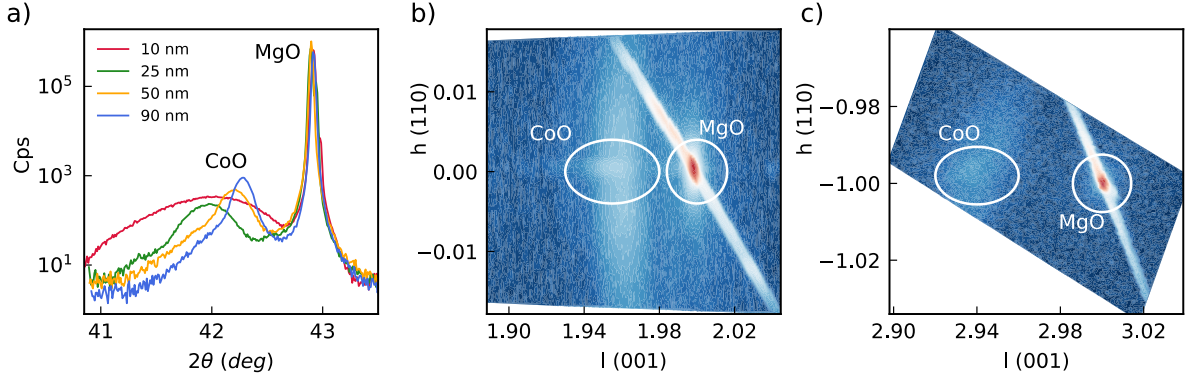

*Fig. S1: (a) 2θ-ω XRD scans showing the (001) alignment of the CoO(t)/Pt(2) films on MgO(001). (b) Symmetric reciprocal space mapping data around the MgO(002) peak indicating a larger out-of-plane lattice parameter for CoO, consistent with the XRD scans. (c) Antisymmetric RSM around the MgO(113) peak, showing a very similar h-value for CoO and MgO, stemming from the correspondence of the in-plane lattice constants between the film and substrate.*

## MAGNETIC CHARACTERIZATION

We next compare the magnetic properties of a MgO(0.5 mm)//CoO(50 nm) thin film and a MgO(0.5 mm) bare substrate, by SQUID magnetometry, to identify possible contributions from ferromagnetically ordered spins. The sample area was approximately 5x10 mm$^2$ in both cases and the substrates came from the same batch and underwent the same cutting procedure with a wire saw. The raw SQUID measurement shows a large diamagnetic background. After subtraction of the diamagnetic background and normalization of the signal to the substrate volume, we find a very small (<1 A/m) non-linear component, as shown in Fig. S2. Since the signal is slightly larger in the bare substrate compared to the substrate where a thin film was deposited on top, we conclude that this small non-linear component is not due to the CoO thin film, but either due to impurities in the substrate or to the cutting procedure we used (involving a metallic wire). We did not see any evidence for ferromagnetic components in the CoO layer also in synchrotron-based x-ray magnetic linear dichroism-photoemission electron microscopy measurements (XMLD-PEEM, not shown) and in the electrical measurements, as we discuss below, so that it is reasonable to consider that our CoO thin films are collinear compensated antiferromagnets, as expected based on the bulk magnetic order without significant other magnetic ordering or superparamagnetic contributions.

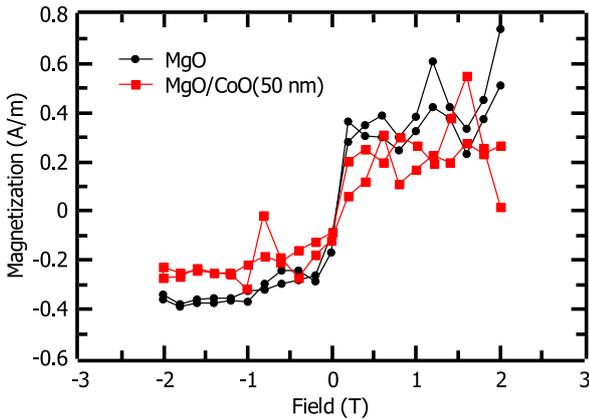

*Fig. S2: SQUID measurements of the MgO substrate and a MgO/CoO(50 nm) sample, after subtraction of the diamagnetic linear component. While a very small non-linear component is seen from the substrate or the cutting procedure, we do not see any significant contributions to the SQUID measurements coming from the CoO thin film.*

## MACROSPIN MODEL OF THE SPIN-FLOP IN AN ANTIFERROMAGNET WITH BIAXIAL ANISOTROPY

In this section we consider a macro-spin model to explain the angular dependence of the spin Hall magnetoresistance (SMR) signal and to estimate the value of the spin flop field. We assume that the magnetic anisotropy of the CoO has tetragonal symmetry and we thus model the energy density as

$$w_{an} = -H_{an}M_s(n_x^4 + n_y^4) + H_{out}M_s n_z^2,$$

where the constants $H_{an}, H_{out} > 0$, parametrise the in-plane and out-of-plane anisotropy fields, respectively, $M_s/2$ is a sublattice magnetization. The Zeeman energy of an antiferromagnet in the presence of an external magnetic field **H** is represented in a standard way as:

$$w_{Zee} = -\frac{M_s}{2H_{ex}}(\mathbf{H} \times \mathbf{n})^2,$$

where $H_{ex}$ is the value of the exchange field responsible for the antiparallel alignment of the two magnetic sublattices.

We consider an external magnetic field applied parallel to the plane of the sample, which keeps the Néel vector in plane, so that $n_z = 0$, consistent with the in-plane anisotropy of the film. In this case, the orientation of the Néel vector is unambiguously described by the angle $\varphi$ with respect to one of the easy axes. The equilibrium orientation of **n** corresponds to the minima of the energy $(w_{an} + w_{Zee})(\varphi)$.

Once the equilibrium value of $\varphi_{eq}$ is known, the value of the SMR is calculated as $R_{transv} = -\Delta R(\mathbf{H})n_x n_y = -0.5\Delta R(\mathbf{H})\sin(2\varphi_{eq})$, where the fitting coefficient $\Delta R(\mathbf{H})$ can depend on the orientation and magnitude of the magnetic field. We additionally comment that, in the transverse resistance data of Fig. 1 of the main text, we observe in addition to the current-induced spin-flop signal, that has the symmetry and sign of a negative SMR, a weak signal with the symmetry of a positive SMR (parabolic background in Fig. 1c,d). A signal with the same symmetry has been already reported in Pt/CoO/Pt trilayers [1], and was explained based on large canting angles of the CoO spins. Even if this signal might indicate the presence of a surface magnetization [2], CoO exhibits spins ferromagnetically aligned in (111) planes, while neighboring planes are aligned antiferromagnetically along the [111] direction. This implies that the spins at the CoO(001) interface are aligned in a checkerboard fashion and the interface with the Pt is expected to be fully compensated [3]. As discussed above, a ferromagnetic signal coming from an interfacial ferromagnetic layer was not revealed by SQUID magnetometry and XMLD-PEEM within our experimental sensitivity. Moreover, this parabolic signal decreases with increasing temperature, does not vanish at the CoO Néel temperature and its field dependence is not influenced by the Néel vector orientation, so that it does not seem coupled to the CoO antiferromagnetic spins. Finally, we have not seen in the electrical measurements an effect of the sign of the magnetic field, nor the presence of a hysteresis loop in the uniaxial MR scans, which are the typical features of signals coming from ferromagnetic layers. To show this, in Fig. S3 we plot the same data from Fig. 1c as a function of the modulus of the field, so that one can directly compare between positive and negative fields. Together, these results suggest that the parabolic signal is not related to the presence of uncompensated magnetic moments at the interface, but rather to a field-induced mechanism occurring at the interface or in the Pt. In this paper, we will treat the parabolic signal as a magnetoresistive background, as it does not influence switching experiments performed at constant field and does not affect our conclusions on the efficiency of the torques and on the comparison between spin-flops induced by fields and currents.

**Methodological note**: in this paper, we consider the transverse resistance ($R_{transv} = [(V(I_{meas}^+)/I_{meas}^+) + (V(I_{meas}^-)/I_{meas}^-)]/2$) as the average of positive and negative currents. The transverse offset resistance signal ($R_{transv,off} = [(V(I_{meas}^+)/I_{meas}^+) - (V(I_{meas}^-)/I_{meas}^-)]/2$), that one can also consider, does not depend on the sign of the current and can have many different origins (electronic offsets, Seebeck effect, spin Seebeck effect, etc.), so that it is not easy to interpret. In our measurements, the variation of the transverse offset signal ($R_{transv,off}$) at $j_{meas} \sim 5 \times 10^9$ A $m^{-2}$ is always one order of magnitude smaller than the "average" resistance signal ($R_{transv}$) in Hall bars. We plot in Fig. S3b one example of the transverse "offset" resistance to show that it does not provide useful information for the present study, in the relevant experimental conditions, while we proceed below to show how the macrospin model explains the transverse resistance data.

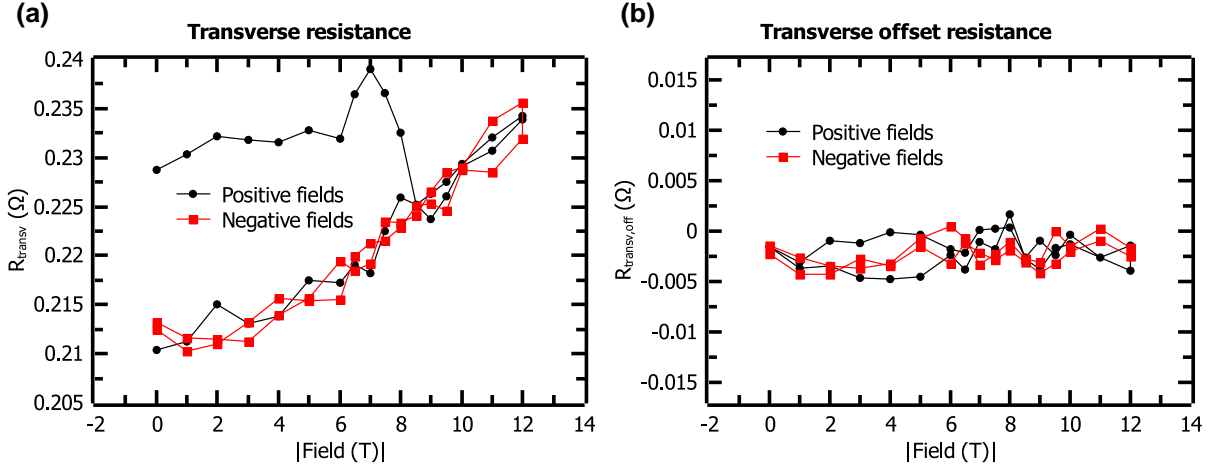

*Fig. S3: (a) Comparison between transverse resistance data at positive and negative fields from Fig 1c of the main text. As expected for antiferromagnetic materials, the sign of the magnetic field does not play a role and no hysteresis loop is seen in uniaxial scans. (b) Transverse offset resistance data acquired at the same time as in panel S3a.*

**Uniaxial scans:** The value of the macrospin spin-flop field can be estimated from the uniaxial scans when the magnetic field is applied along one of the easy axes. In this case, from the minimization of the energy $(w_{an} + w_{Zee})(\varphi)$ we find two stable states below the spin-flop field $H \leq H_{SF} = \sqrt{H_{ex}H_{an}}$ corresponding to the alignment along two easy direction: metastable state with $\boldsymbol{n}||\boldsymbol{H}$ and stable state with $\boldsymbol{n} \perp \boldsymbol{H}$. Above the macrospin spin-flop field, $H > H_{SF}$, only one state with $\boldsymbol{n} \perp \boldsymbol{H}$ is stable. Fig. S4a shows the calculated and measured field dependencies of the SMR for the field scan starting from the metastable state $\boldsymbol{n}||\boldsymbol{H}$. After the first crossing of the spin-flop field, the Néel vector flops to the stable state $\boldsymbol{n} \perp \boldsymbol{H}$, which is then stable and cannot be changed by further variation of field value, but only by a change of the field orientation, as shown in Fig. 1c,d of the main text.

**Angular dependence:** Magnetic fields, applied at a generic angle with respect to the easy axis, induce rotation of the Néel vector toward the direction perpendicular to $\boldsymbol{H}$. The minimization of $(w_{an} + w_{Zee})(\varphi)$ shows that, above a threshold magnetic field and below the macrospin spin-flop field $H \leq H_{SF}$, two equilibrium states are stable in certain ranges of angles. Thus, angular scans at a fixed field value can induce step-like reorientation the Néel vector in the points where one of the states loses stability, which experimentally occurs below the theoretical macrospin spin-flop field. Fig. S4b shows the calculated and measured field dependencies of the SMR for the ADMR scan starting from the stable state $\boldsymbol{n} \perp \boldsymbol{H}$.

**Remarks:** By the macrospin antiferromagnetic model presented above, we can qualitatively explain both the uniaxial magnetoresistance (MR) scans and the hysteresis loops observed in the angular-dependent magnetoresistance (ADMR) measurements. This confirms that the magnetic anisotropy is biaxial in-plane. However, we had to artificially consider different macrospin spin-flop fields for the MR (8 T) and ADMR (12.5 T) measurements. This discrepancy is related to the existence of a high-energy metastable state, that is important in the uniaxial scans at fields below the macrospin spin-flop field, when the field is applied along the easy axis parallel to $\boldsymbol{n}$, but this metastable state is never accessed in the ADMR hysteresis loops. This discrepancy can be explained by considering that the evolution of the system is more complex than the macrospin model described here, as thermally activated processes allow for the evolution of the system from the high-energy metastable state toward the lower energy state, and one should furthermore consider the role of domains and domain walls that would decrease the barrier of the metastable state, which we do not do here.

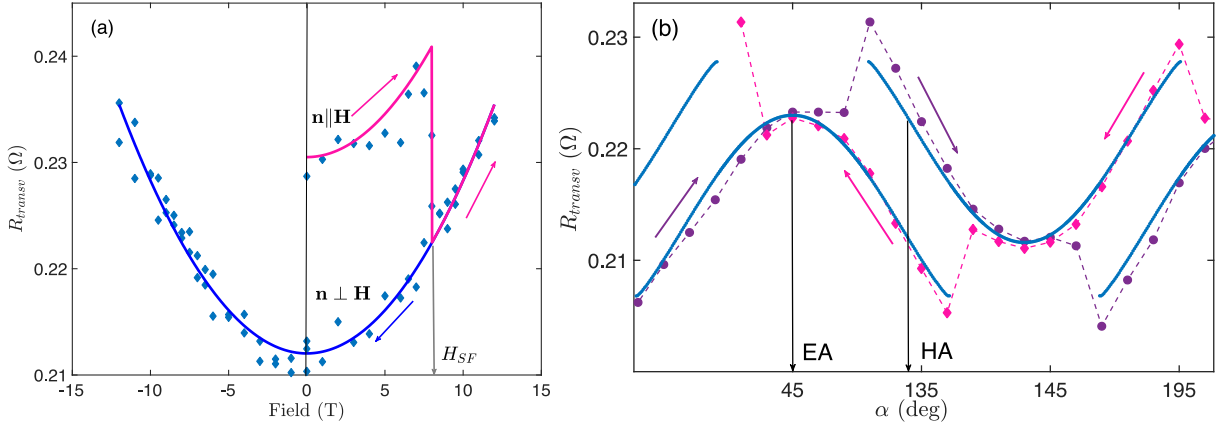

*Fig. S4: Reorientation of the Néel vector probed by SMR. (a) Uniaxial magnetoresistance measurements in the presence of a magnetic field applied along the easy axis. The solid line shows the calculated dependence (assuming $H_{SF} = 8\ T$) on top of the quadratic ($\propto H^2$) background, while the points show the experimental data. (b) Angular dependence of the SMR in the presence of a field H=9 T. The solid line shows the calculated dependence (assuming $H_{SF} = 12.5\ T$) on top on the $\cos(2\alpha)$ background. The vertical lines show the position of the easy (EA) and hard (HA) magnetic axes.*

## COMPARISON BETWEEN FIELD-INDUCED AND CURRENT-INDUCED SPIN-FLOP

To estimate the fraction of domains that can be switched by current pulses, we compare the current-induced switching shown in Fig. 2 of the main text and the field-induced spin-flop transition in the same Hall star device patterned on a CoO 5 nm/Pt 2 nm sample. The field-induced spin-flop is obtained by alternating the field in the α = 45° and α = 135° directions (i.e. before each measurement an orthogonal 11 T field was applied) and the results are shown here in Fig. S5. One can see that the amplitude of the switching is $\Delta R_t = 9.7 \pm 0.5\ m\Omega$ in the case of current-induced switching at 200 K for 5x 1 ms pulses of current 23 mA (Fig. 2), while it is $\Delta R_t = 10.4 \pm 0.5\ m\Omega$ in the case of field-induced switching (Fig. S5). These two amplitude values are compatible within each other, indicating that currents and fields can both manipulate effectively the spin system in the same manner. One can also note that the shape of the magnetoresistance curves after current-induced or field-induced switching are identical within the noise for this saturated pulse current, confirming unambiguously the occurrence of current-induced magnetic switching.

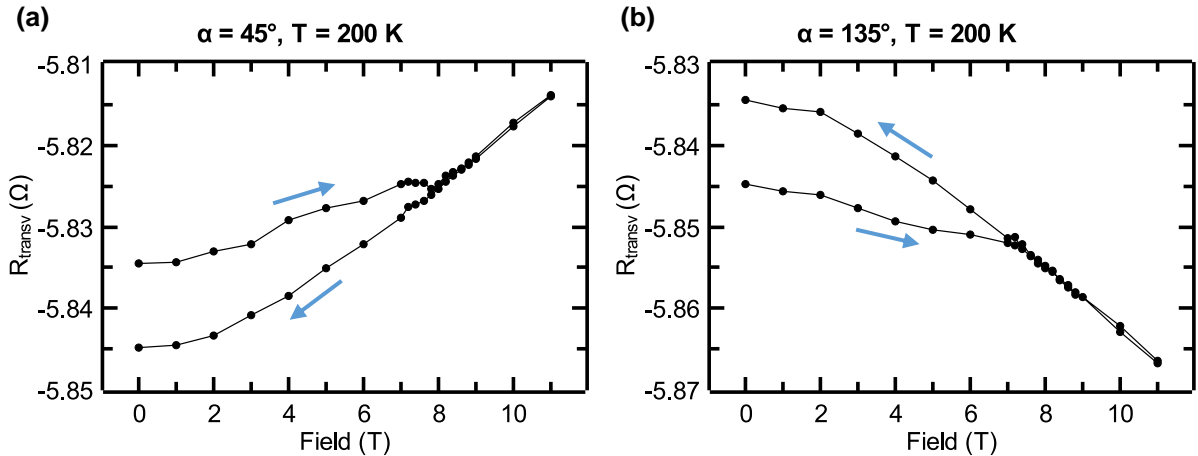

*Fig. S5: Uniaxial MR measurements after the field-induced spin-flop of the Néel vector along the two different easy axes, to be compared with the current induced switching at saturation. The transverse resistance variation by field-induced spin-flop and current-induced spin-flop discussed in the main text are compatible within each other.*

# REPRODUCIBILITY OF THE SWITCHING

To check the reproducibility of the switching we alternated 2x +45° and 2x -45° pulses many times in a Hall star device in a CoO sample 5 nm thick at 250 K. We applied 1 ms long pulses at 20 mA (corresponding to $1 \times 10^{12}$ A m$^{-2}$, which saturates the switching at that temperature, see Fig. S7 and related discussion) and measured the transverse resistance as shown in Fig. S6. One can see reproducible reversible switching for almost 350 times (one switching every two pulses). The states can be clearly distinguished and the device was still working after the end of the sequence.

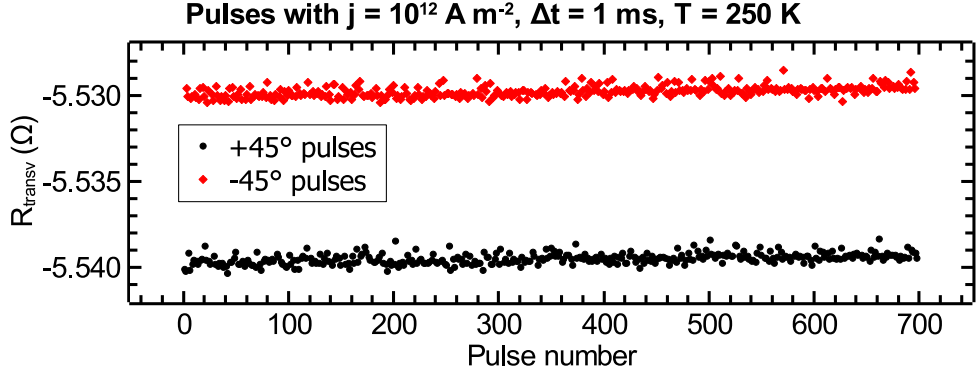

*Fig. S6: Reproducibility of the current-induced magnetic switching at 250 K. The black points are acquired after +45° pulses, the red points after -45° pulses.*

# CURRENT-FIELD EQUIVALENCE FROM THE SPIN-FLOP FIELD OF THE SWITCHED STATES

To verify the possibility to tune the switching and if it allows for a multi-level final state, we measured the uniaxial MR after having applied current pulses of different amplitude, as shown in Fig. S7. Note that, in this particular sample and device, the spin-flop is a smooth transition as a function of the field. One can see that the transverse resistance variation and the spin-flop field of the switched state change as a function of the pulse amplitude, indicating that the system entails a distribution of domains with different magnetic anisotropy and pinning with slightly different thresholds. By the difference of spin-flop field $\Delta H_{sf} = 0.5\ T$ between the states obtained after the pulses at 20 and 23 mA, we can obtain a first estimation of the order of magnitude of the current field equivalence, of 3x $10^{-12}$ T A$^{-1}$ m$^2$. This is smaller than the value obtained by the method based on pulses applied in the presence of magnetic field described in the main text (Fig. 3 and related discussion). However, this second method described here tends to underestimate the current-to-field ratio, as the spin-flop field of the domains is limited in range and cannot increase above a maximum value. In Fig. S7b we apply pulses 1-5 at 45° and pulses 6-10 at -45°, according to the convention shown in Fig. 2, showing that the switching is reversible and entails a threshold and a saturation, thus further confirming the magnetic origin of the switching.

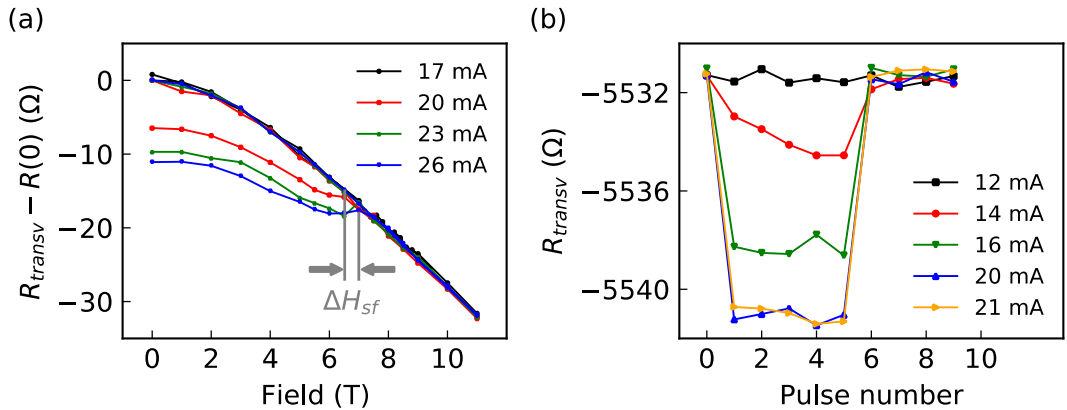

*Fig. S7: (a) Pulse current dependence of the Néel order switching in CoO 5 nm/Pt 2 nm at 200 K. The spin-flop field increases with the increasing pulse current. (b)*

*Current-induced reversible switching at 250 K from threshold to saturation. Pulses 1-5 were applied at 45°, while pulses 6-10 were applied at -45°. Note that the set/reset operations can be achieved by a current alone and no applied magnetic field is necessary to induce deterministic switching.*

## DERIVATION OF THE CURRENT-FIELD EQUIVALENCE IN THE THERMOMAGNETOELASTIC TORQUE MODEL

In the present section we derive the expression for the current-field equivalence in the thermomagnetoelastic mechanism related to Joule heating. It results from the combined effect of thermal expansion and magnetoelasticity [4]. According to this model, the inhomogeneous distribution of the current density $j = j_{pulse} f(\mathbf{r})$ in the Pt electrode creates a corresponding inhomogeneous temperature profile within the film plane (where $j_0$ is the average current density far from the cross). This, in turn, induces local volume expansions $u^v_{xx}(\mathbf{r}) = u^v_{yy}(\mathbf{r}) = u^v_{zz}(\mathbf{r}) = \alpha T(\mathbf{r})/3$ in CoO ($\alpha$ is the thermal expansion coefficient) which maps the temperature distribution $T(\mathbf{r})$. Obviously, the value of thermal expansion is larger in the hotter parts and smaller in the colder parts of the sample. Hence, in the regions with nonzero temperature gradient, the strains $u^v_{jj}(\mathbf{r})$ are incompatible along the isotherms (Fig. S8). These incompatibilities create additional stresses which are relaxed due to the formation of additional shear strains $u^{rel}_{jk}(\mathbf{r})$, whose geometry depends on orientation and distribution of the temperature gradient. The strains along the direction of temperature gradient are tensile at the hotter side and are compressive at the colder side, as explained in Fig. S9a. Through the magnetoelastic interactions, parametrized with the magnetoelastic constant $\lambda$, these additional strains contribute into magnetic anisotropy as $w_{me} = \lambda u^{rel}_{jk} n_j n_k$ and can remove the degeneracy of the orthogonal states. Due to the non-locality of the elastic interactions, the effective contribution into the magnetic anisotropy depends on the distribution of the current density gradients, with respect to the observation point, and it is related with the direction of $\boldsymbol{j}$ only indirectly, through the convolution of $\nabla f(\mathbf{r})$ with the kernel $K_j(\mathbf{r} - \mathbf{r}')$, whose structure is defined by the elastic and magnetoelastic properties of the sample:

$$w_{me}(\mathbf{r}) = j_0^2 \frac{\alpha \lambda}{\kappa \sigma} n_j n_k \int dV' K_j(\mathbf{r} - \mathbf{r}') \partial_k f(\mathbf{r}'). \quad (S1)$$

Here $\sigma$ is the conductivity of Pt, $\nu$ and $\kappa$ are the Poisson ratio and thermal conductivity of CoO, $\alpha$ is the thermal expansion coefficient. By comparing this equation with Eq. (2) of the main text we conclude that the effect of current-induced strains $u^{rel}_{jk}(\mathbf{r})$ is equivalent to the effect of an inhomogeneous magnetic field, whose orientation is defined by the integral $A = \int dV' K_j(\mathbf{r} - \mathbf{r}') \partial_k f(\mathbf{r}')$ and the sign of the magnetoelastic constant $\lambda$. For our experimental geometry, in the center of the cross the effective field is oriented perpendicular to the current, assuming that $\lambda < 0$ (see Fig. S9b).

The effective expression for the magnetoelastic energy in this region can be then approximated as

$$w_{me} = A \frac{\alpha |\lambda|}{\kappa \sigma} (\boldsymbol{n} \cdot \boldsymbol{j}_{pulse})^2, \quad (S2)$$

By comparing Eq. (S2) with the expression for the Zeeman energy (Eq. (2) of the main text) we extract the value of the effective field

$$H_{me} = j_{pulse} \sqrt{\frac{\alpha |\lambda|}{\kappa \sigma M_s} A H_{ex}}. \quad (S3)$$

One can notice that the effective field scales with the current density $j_{pulse}$. Further comparison shows that, depending on the mutual orientation between $\boldsymbol{H}$ and $\boldsymbol{j}_{pulse}$, the effects of the magnetic field and current can compete (if $\boldsymbol{j}_{pulse} \perp \boldsymbol{H}$) or sum up (if $\boldsymbol{j}_{pulse} \| \boldsymbol{H}$), thus increasing or reducing the threshold value of the current as a function of the applied field.

Note that this model determines the final state of the switching in the center of the cross, while determining the spatial extension of the switched area requires spatially resolved simulations of the strain, so that the center of the cross and the arms could exhibit the same switching behavior.

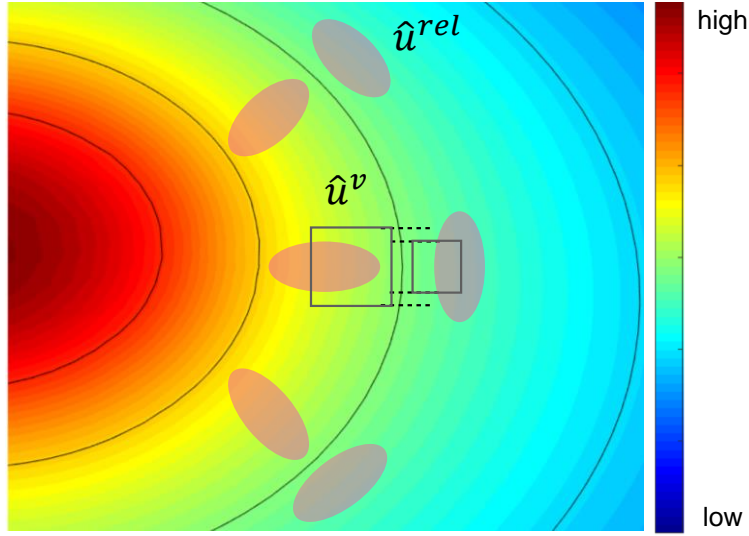

*Fig. S8 (Color online). Cartoon of the thermo-magneto-elastic effect. The spontaneous volume expansion $\hat{u}^v$ (squares) is larger in hotter region and is incompatible along isotherms, as shown with dashed lines. The orientation of the principal axes of shear strains $\hat{u}^{rel}$ (shown as ellipses), which compensate the incompatibilities, depends on the temperature distribution (color code).*

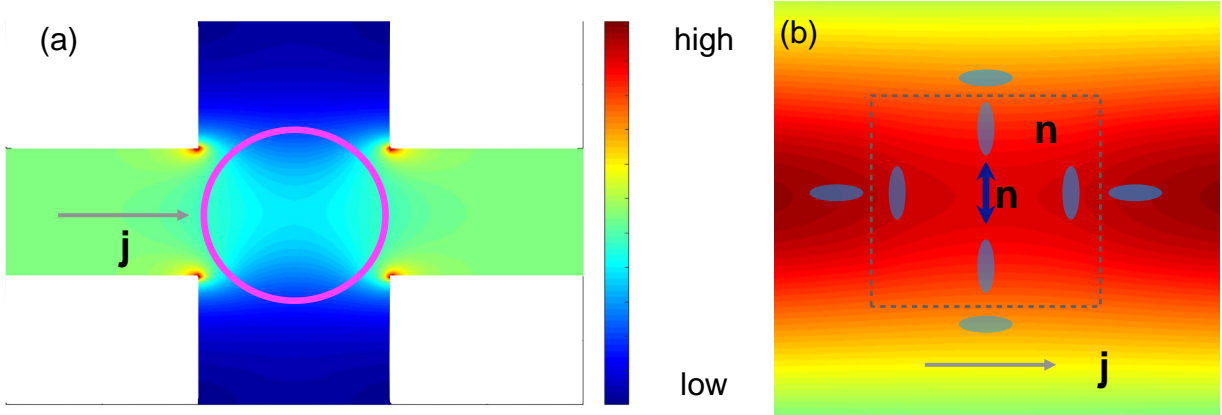

*Fig. S9 (Color online). Preferable direction of the Néel vector in the center of cross due to current-induced thermomagnetoelastic effect (cartoon). (a) Distribution of the current density |**j**| (color code) in the Pt electrode. Due to the geometry of the device and in the presence of "straight" pulses, the current density has a saddle point in the center of the cross (minimum along the horizontal direction and maximum in the vertical one). (b) The temperature distribution (color code) created by Joule heating in the center of the cross favors the extension $\hat{u}^{rel}$ (ellipses) and alignment of the Néel vector **n** perpendicular to the current density **j**.*

## REFERENCES SUPPLEMENTARY